\documentstyle[epsfig]{mn}
\begin{document}

\def\ltsima{$\; \buildrel < \over \sim \;$}
\def\simlt{\lower.5ex\hbox{\ltsima}}
\def\gtsima{$\; \buildrel > \over \sim \;$}
\def\simgt{\lower.5ex\hbox{\gtsima}}
\def\ls{{_<\atop^{\sim}}}
\def\lax{{_<\atop^{\sim}}}
\def\gs{{_>\atop^{\sim}}}
\def\gax{{_>\atop^{\sim}}}
\def\cgs{ ${\rm erg~cm}^{-2}~{\rm s}^{-1}$ } 

\title[Faint AGN and the X-ray background]
{The contribution of faint AGN to the hard X-ray background}

\author[F. Fiore et al.]{F. Fiore$^{1,2,3}$,  
F. La Franca$^4$, P. Giommi$^1$,  M. Elvis$^3$, G. Matt$^4$, 
A. Comastri$^5$, \\ ~ \\
{\LARGE S. Molendi$^6$, and I. Gioia$^{7,8}$}\\ ~ \\
$^1$ BeppoSAX Science Data Center, Via Corcolle 19, I--00131 Roma, Italy\\
$^2$ Osservatorio Astronomico di Roma, Via Frascati 33,
I-00044 Monteporzio, Italy\\
$^3$ Harvard-Smithsonian Center of Astrophysics, 60 Garden Street, 
Cambridge MA 02138 USA\\
$^4$ Dipartimento di Fisica, Universit\`a degli Studi ``Roma Tre",
Via della Vasca Navale 84, I--00146 Roma, Italy \\
$^5$ Osservatorio Astronomico di Bologna, via Ranzani 1, I40127
Bologna, Italy \\
$^6$ IFCTR/CNR, via Bassini 15, Milano, I20133, Italy\\
$^7$ IRA/CNR, via Gobetti 101, I40129 Bologna, Italy\\
$^8$ Institute for Astronomy, Hawaii, 96822 USA
}

\maketitle
\begin{abstract}

Hard X-ray selection is the most efficient way to discriminate between
accretion-powered sources, such as AGN, from sources dominated by
starlight.  Hard X-rays are also less affected than other bands by
obscuration.  We have then carried out the BeppoSAX High Energy Large
Area Survey (HELLAS) in the largely unexplored 5-10 keV band, finding
180 sources in $\sim 50$ deg$^2$ of sky with flux $\gs5\times10^{-14}$
\cgs~.  After correction for the non uniform sky coverage this
corresponds to resolving about 30 \% of the hard Cosmic X-ray
Background (XRB). Here we report on a first optical spectroscopic
identification campaign, finding 12 AGN out of 14 X-ray error-boxes
studied. Seven AGN show evidence for obscuration in X-ray and optical
bands, a fraction higher than in previous ROSAT or ASCA-ROSAT surveys
(at a 95-99 \% and 90 \% confidence level respectively), thus supporting
the scenario in which a significant fraction of the XRB is made by
obscured AGN.

\end{abstract}


\section{Introduction}

While it is now clear that the Cosmic X-ray Background (XRB) is made
by the superposition of many discrete sources, and that the soft
(0.5--2 keV) XRB is mostly produced by AGN (Hasinger et al. 1998,
Schmidt et al. 1998), the nature of the sources making the
energetically dominant hard (2--50 keV) XRB is still largely unknown.
This is due to the lack, until recent years, of sensitive imaging
instruments above 2--3 keV. Thanks to ASCA and BeppoSAX, the 2-10 keV
band is now accessible to surveys (Boyle et al. 1998, Ueda et
al. 1998, Giommi et al. 1998, Giommi, Fiore \& Perri 1998, Fiore et
al. 1998a,b).  The hard X-ray sky poses major problems: no known large
class of sources has an emission spectrum matching the kT$\sim40$~keV
thermal--like spectrum of the XRB; fluctuation analyses of Ginga data
(Warwick \& Stewart 1989), and ASCA (Georgantopoulos et al. 1997,
Cagnoni et al.  1998) and BeppoSAX (Giommi et al. 1998, Giommi, Fiore
\& Perri 1998) 2--10 keV source counts, require 2--3 times as many
sources as expected from ROSAT counts, if they have the steep power
law spectrum typical of ROSAT sources ($1<\alpha_E<2$, $F(E)\propto
E^{-\alpha_E}$).  The leading suggestion to reconcile the source
counts, based both on theoretical grounds (Setti \& Woltjer 1989,
Madau et al. 1994, Matt \& Fabian 1994, Comastri et al. 1995) and high
energy observations of bright nearby AGN (Zdziarski et al 1995, Smith
\& Done 1996), is that heavily obscured AGN, emerging strongly at high
energies, are the main contributors to the hard XRB. (These AGN are
probably not completely invisible at low X-ray energies, i.e. $\sim 1$
keV, because even if the nuclear emission is completely blocked,
different components, like starburts, optically thin gas or scattering
of the nuclear radiation, may still be detectable at the 1\%--10\%
level, Giommi, Fiore \& Perri 1998, Schachter et al. 1998).  It is
crucial to test these suggestions observationally in a band where the
nucleus is directly visible and down to fluxes where the bulk of the
hard XRB is produced.  Hard X-ray selection would also help in
distinguishing between different scenarios. For instance, if the bulk
of the XRB is made by obscured luminous `quasar~2', following the
evolution of type 1 AGN, the peak of their activity should be at
z=2--3 and then decrease quickly toward lower redshifts.  On the other
hand, the XRB could be mostly made by a large population of less
luminous/active obscured AGN, spread in a broader interval of
redshifts. The solution of these problems may have impact also on AGN
unification schemes.

The BeppoSAX MECS (Boella et al. 1997a,b) provides a good opportunity
to investigate the hard X-ray sky, thanks to its good sensitivity
above 5 keV (5-10 keV flux limit of $\sim0.002$ mCrab in 100 ks), and
improved point spread function (PSF). We have thus carried out the
High Energy LLarge Area Survey (HELLAS). The survey has been performed
in the hard 5-10 keV band because this is the band closest to the
maximum of the XRB energy density that is reachable with the current
imaging X-ray telescopes. Including the softer 1.5-5 keV range would
only increase the background for faint, heavily absorbed sources, thus
reducing the chances of their detection.  Secondly, the BeppoSAX MECS
PSF improves with energy, and in the 5-10 keV band provides error
circles of 1~arcmin, 95\% confidence radius, small enough to allow the
optical identification of the X-ray sources.  The HELLAS survey has so
far cataloged 180 sources (Fiore et al. 1999, in preparation).  The
sky coverage is $\sim2-50$ square degrees at
$F_{5-10keV}=5-50\times10^{-14}$ \cgs respectively.  At the fainter
limit we find between 16 and 20 sources deg$^{-2}$ (Fiore et
al. 1998a,b, Giommi, Fiore \& Perri 1998, Comastri 1998), implying
that about 30 \% of the XRB is resolved. Cross-correlations of the
HELLAS sample with catalogs of cosmic sources provides 18 coincidences
(7 radio-loud AGN, 6 radio-quiet AGN, 3 clusters of galaxies, 1 CV and
1 normal star), suggesting that most of the HELLAS sources are
AGN. However, the 13 AGN were discovered in radio, soft X-ray, and
optical surveys, and so this sample is biased against highly obscured
sources. To remove this bias we have started a program to
spectroscopically identify the rest of the HELLAS sample. In this
letter we report on the first results of this program. The sources
studied have not been selected according to their X-ray or optical
properties but solely because of visibility during our observing runs.
As a result, they can be considered representative of the whole HELLAS
sample.

\section{Optical Identifications}

We carried out spectroscopic identification of optical candidates for
the HELLAS sources using the RC spectrograph (RCSP) at the Kitt Peak
4m telescope for 9 sources, the FAST spectrograph at the Whipple 60''
telescope for 1 source, the Hawaii 88'' for 2 sources and EFOSC2 at
the ESO 3.6m for 2 sources. Long-slit spectra have been obtained in
the 3450-8500 $\AA$~range (RCSP) or 3800-8000 $\AA$~range (FAST, 88'',
EFOSC2), with a resolution between 7 and 16 \AA. Between 1 and 7
optical candidates were identified on APM or Cosmos scans of the E (R)
POSS plates down to R$\sim 20$m, within a conservative (95 \%)
error-circle radius of 60$''$.  The complete set of spectra will be
presented elsewhere (La Franca et al. 1999 in preparation).  We
identified the X-ray source with the more plausible object in the
error box (if any), i.e. either sources with high hard X-ray to
optical ratio (like AGN, Galactic binaries, clusters and groups of
galaxies) or bright galaxies and stars.  Unlike previous
identification campaigns of ASCA sources, we decided not to take
advantage of correlations with ROSAT source catalogs or Radio
catalogs, to avoid selection biases against obscured objects.
Instead, our basic strategy is to obtain spectra of all candidates
down to the chosen R=20 magnitude limit. However, in 4 cases a few
faint candidates still remain to be observed in each error-box. Their
magnitude is fainter than that of the candidate identified as an AGN,
and therefore the identification of the X-ray source with the AGN is
rather secure.  In twelve cases we obtained a unique identification
(only one plausible candidate in the error-box). In one case we found
2 sources in the error-circles which could contribute to the detected
hard X-ray flux (designated A,B in Table 1). In one case none of the 6
objects observed with $17<R<20$ were good candidates for being the
X-ray source.  All fields but one (0122.1-5845) have been observed
with the VLA as part of the NVSS survey. No radio source brighter than
3 mJy is found within any of the HELLAS error-boxes. Given the
sources' optical magnitude this implies that all objects can be
considered as radio-quiet. Table 1 gives, for each of the identified
sources, the position of the optical counterpart and its distance from
the X-ray centroid in arcsec, its redshift, R and B magnitudes,
optical and X-ray luminosities and the classification.

\begin{table*}
\label{pos}
\caption{\bf HELLAS sources}
\begin{tabular}{lcccccccccccc}

1SAXJ            & $F_X^a$ & log$N_H^b$ & $L_X^c$ &  Ra$^d$ & Dec$^d$
& R & B & D$^e$ & M$_V$ & z & Class$^f$ & PSPC$^g$ \\

0027.1-1926$^h$ & 1.7$\pm$0.3 & $<21.5$  & 43.7 & 00 27 09.8 & -19 26 13 
& 17 & 18 & 26 & -23.9 & 0.227 & B & --\\

0045.7-2515 & 3.5$\pm$1.0  & 22.6$\pm$0.5  & 43.3 & 00 45 46.3 & -25 15 50
& 17.4 & 17.9 & 58 & -21.3 & 0.111 & 1.9 & det. \\ 

0122.1-5845 & 1.5$\pm$0.4 & 23$\pm$1  & 43.0 & 01 21 56.9 & -58 44 41
& 18.4 & -- & 60 &  -21.0 & 0.118 & 1.9 & lim. \\

1032.3+5051 & 2.2$\pm$0.8 & $<21.2$ & 43.5 & 10 32 16.1 & 50 51 21
& 15.5 & 16.8 & 18 & -24.8 & 0.174 & B & --\\

1117.8+4018 & 1.3$\pm$0.5  & $22.7\pm0.5$ & 45.4 & 11 17 48.7 & 40 17 54
& 19.9 & -- & 22 & -25.9 & 1.274  & B & det.\\

1118.2+4028 & 0.9$\pm$0.3  & $<22.1$ & 43.9 & 11 18 13.9 & 40  28  38
& 18.3 & 20.9 & 23 & -24.0 & 0.387 & R & det.\\

1118.8+4026A$^{i}$& 1.2$\pm$0.4 & $<22.5$ & 45.1 & 11 18 48.7 & 40 26 48
& 18.0 & 18.4 & 60 & -27.4 & 1.129  & B & det.\\

1118.8+4026B$^{i}$& &  $<22.4$ &  &  11 18 47.9 & 40  26  44
& 20.0 & 20.7 & 60 & -24.4 & 0.888 & B & det.\\

1218.9+2958 & 2.4$\pm$0.7  & 23.1$\pm$0.4  & 43.6 & 12 18 52.5 & 29 59 01
& 18.6 & 21.0 & 53 & -21.6 & 0.176 & 1.9 & det.\\

1353.9+1820& 8.5$\pm$2.3  & $22.8_{-1.2}^{+0.3}$ & 44.2 & 13 53 54.4 
& 18  20  16 & 17.1 & 20.0 & 31 & -23.7 & 0.217 & R & lim. \\

1519.5+6535$^l$ &10.8$\pm$2.2  & 23.2$\pm$0.1$^m$ & 43.2 & 15 19 33.7 
& 65 35 58 & 15.5 & 16.8 & 29 & -21.7  & 0.044 & 1.9 & lim.\\

1528.8+1939 & 1.4$\pm0.3$ &  $<22.6$ & 43.3 & 15 28 49.0 & 19 38 52
& 19.2 & 20.7 & 35 & -21.1 & 0.178 & L & --\\

1634.3+5945& 0.8$\pm$0.1  & 22.8$\pm$0.3  & 43.7 & 16 34 12.9 & 59 45 04
& 19.0 & 21.1 & 18 & -22.5 & 0.341 & 1.8 & -- \\    

2226.5+2111 & 4.2$\pm$1.5  & $<21.5$ & 44.2 & 22 26 31.5 & 21 11 35
& 16.8 & 17.5 & 8 & -23.7 & 0.260 &  B & det.\\

1134.7+7024 & 3.7$\pm$1.4 & 23$\pm1^n$ & -- & -- & -- & -- & -- & -- 
& -- & -- & -- & lim. \\

\end{tabular}

$^a$ 5-10 keV flux in $10^{-13}$ \cgs;
$^b$ $N_H$ values in log cm$^{-2}$ are estimated from the 
MECS 5-10 keV/1.3-5 keV or from the MECS 5-10 keV/PSPC 0.5-2 keV hardness 
ratios, when ROSAT PSPC data are available, and from a long ASCA
observation in one case, assuming $\alpha_E=0.8$;
$^c$ log 5-10 keV luminosity in erg s$^{-1}$
calculated assuming a power law model with 
$\alpha_E=0.8$ and no absorption, for log$N_H$=23 the unobscured
intrinsic 5-10 keV luminosity is higher by $\sim20$ \%
and a 10 \%  at z=0 and 0.4 respectively.
$H_0 =  50.0$ and  $q_0 = 0.0$ have been assumed; 
$^d$ J2000;
$^e$ distance between optical and X-ray position in arcsec;
$^f$ B=Blue broad line QSO, R=Red broad line QSO, 1.8-1.9= intermediate
type 1.8-1.9 AGN, L=LINER;
$^g$ The optical counterpart lies in all cases  
within the PSPC error-box of 30 arcsec radius;
$^h$ Colafrancesco et al. (1998);
$^i$ 1SAXJ1118.8+4026A is brighter by about 2
magnitudes brighter than 1SAXJ1118.8+4026B in both R and B bands, therefore
it is probably the one giving the largest contribution to the X-ray flux;
$^l$ CGCG319-007;
$^m$ from ASCA data;
$^n$ at z=0;

\end{table*}

\begin{figure}
\centerline{ 
\epsfig{figure=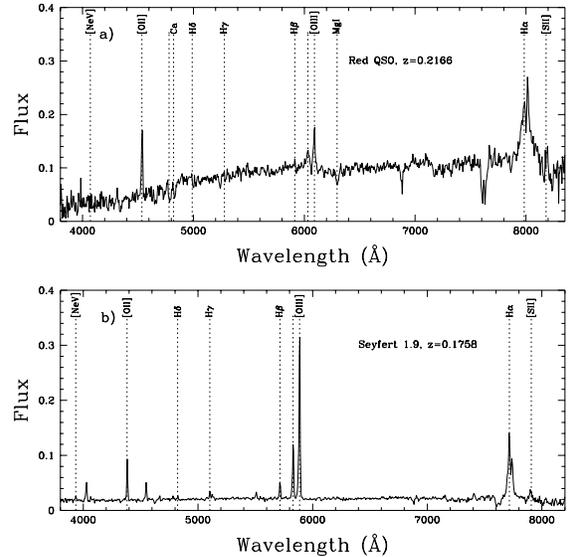, height=8.cm, width=8cm,angle=0}
}
\caption[h]{
The spectra of the red quasar 1SAXJ1353.9+1820 and of the
type 1.9 AGN 1SAXJ1218.9+2958 
}
 
\end{figure}

\begin{table}
\label{lines}
\caption{\bf Emission lines}
\begin{tabular}{lcccc}

1SAXJ    &  [OIII]/H$\beta$ 
& [OII]/H$\beta$ & [SII]/H$\alpha$ 
&  [NeV]$\lambda3426$ \\
& & & & EW in \AA \\

0045.7-2515  &  $>5$ & --  & 0.1   & --   \\
0122.1-5845  &  12   & 3.5 & 0.2   & --   \\
1218.9+2958  &  10.0 & 2.0 & 0.135 & 20.0 \\
1634.3+5945  &   9.0 & 1.1 & --    & 13.8 \\
1519.5+6535  & $>13$ &$>1.7$& 0.3  & --   \\
1528.8+1939  &   1.3 & 3.1 & 0.5   & --   \\

\end{tabular}
 
\end{table}

Broad emission line quasars with a normal blue continuum,
$-1.3<\alpha< +0.1$ ($F(\nu)=\nu^\alpha$, $0.17<$z$<1.28$), were found
in five error-boxes.  Broad emission line quasars with a very `red'
continuum were found in two cases ($\alpha(5000-8000\AA) \sim-3$,
consistent with their B-R color (see Table 1).  Their spectra are
dominated by starlight, based on the detection of the Calcium H and K
and MgI absorption features and the small equivalent width of
H$\alpha$ (90 \AA) observed in 1SAXJ1353.9+1820. In this case we could
estimate a lower limit to the extinction of A$_V\gs4.8$, corresponding
to log$N_H>21.9$ assuming a Galactic dust to gas ratio. So, this
quasar appears `red' both because the spectrum is dominated by
starlight from the host galaxy and because of the large extinction to
the nucleus.  Narrow emission line galaxies were found in 6
error-boxes.  To identify their nature we used the standard line
ratios diagnostics (Tresse et al. 1996, see Table 2).  Five sources
show large [OIII]$\lambda5007$/H$\beta$ ($\sim 10$) and/or large
[SII]$\lambda6725$/H$\alpha$ ratios. Two sources also show strong
[NeV] lines, which unambiguously identify them as narrow line AGN
(Schmidt et al. 1998).  In all five sources H$\alpha$ or H$\beta$ have
faint wings of FWZI 3000--5000 km s$^{-1}$, broader than the typical
value for AGN Narrow Line Region lines. We then classify these sources
as intermediate 1.8-1.9 type AGN (Osterbrook 1981).  In one
error-boxes we found a narrow emission line galaxy, having
[SII]$\lambda6725$/H$\alpha$ and [OII]$\lambda3727$/H$\beta$ ratios
indicating a lower excitation than in Seyfert 1.8-2 galaxies but close
to or higher than the value which separate HII galaxies from LINER
(Tresse et al. 1996).  We tentatively classify it as a LINER.  Figure 1
shows two examples of HELLAS AGN.

In summary, 12 of the 14 HELLAS error-boxes studied contain AGNs and 1
contains a LINER.  We have checked how many of the optical sources lie
within 30 arcsec from PSPC sources in the WGACAT. We found six
coincidences: 0045.7-2515, 1118.2+4028, 1118.8+4026, 1117.8+4018,
1218.9+2958 (also visible in the ASCA GIS field of Mark766), and
2226.5+2111.  Another 4 sources (0122.1-5845, 1353.9+1820,
1519.5+6535, 1134.7+7024) lie in PSPC fields but are not detected in
the WGACAT.

The surface density of quasars at R=20 (the magnitude of the faintest
of our quasars) is 20--40 deg$^{-2}$ (Zitelli et al. 1992) and
therefore the expected number of chance coincidences in our 14
error-boxes is 0.25-0.5.  The surface density of galaxies at R$\ls$19
(the magnitude of the faintest of our Seyfert 1.8--1.9 galaxies and
LINER) is about 500 deg$^{-2}$ (Lilly et al. 1995).  Tresse et
al. (1996) found that between 8 and 17 \% of the galaxies with z$<0.3$
host Seyfert 1.8--2 or LINER nuclei. Hammer et al. (1997) found that
$\sim11 \%$ of the higher redshift and luminosity CFRS galaxies host
type 1.8--2 nuclei.  The surface density of R$\ls19$ type 1.8--2 AGN
and LINER is therefore 45--90 deg$^{-2}$.  The number of their chance
coincidences in the total area covered by 14 error-boxes is so 0.5--1.
We note that for the AGN this is actually an upper limit, since the
distribution of the magnitudes of the type 1.8--1.9 AGN in Table 1 is
different from a typical galaxy logN-logS, and because two of the type
1.8--1.9 AGN are also PSPC sources.  We then conclude that the
identification of the quasars and type 1.8-1.9 AGN is rather secure,
while for the LINER we cannot exclude a chance coincidence.

\section{X-ray and X-ray/optical colours}

Since for the sources in Table 1 we have redshifts, luminosities,
optical spectroscopy and X-ray colours, we can start addressing the
question of which kind of AGN populates the hard X-ray sky.  X-ray
hardness ratios can be used to estimate the X-ray spectrum.  Assuming
an absorbing column equal to the Galactic column along the line of
sight, produces rather unlikely spectral indices for the 1.8-1.9 type
AGN and the `red' quasars ($-0.5<\alpha_E<0$), harder than the
$\alpha_E=0.4$ of the 3-15 keV XRB spectrum and of any known AGN.  On
the other hand, assuming a typical AGN power law spectrum of energy
index $\alpha_E=0.8$ and the source redshift, we obtain the $N_H$
values given in Table 1. (For the unidentified source we assumed z=0
and so the relative $N_H$ is a lower limit to the true absorbing
column.) The distribution of log$N_H$ is plotted in Figure 2, along
with the distribution predicted by the Comastri et al. (1995) syntesis
model in the 5-10 keV band and at the flux limit of our survey. The
two distributions agree quite well. At a first sight the lack of
heavily absorbed Compton thick (log$N_H>24$, i.e.  the nuclear
emission is completely blocked below 10 keV) sources in our survey is
at variance with the recent findings of Maiolino et al. (1998) and
Risaliti et al. (1999).  However it should be noted that their results
have been derived from an optically selected sample of bright, nearby
Seyfert 1.8-2 galaxies, and thus the comparison with our survey is not
straigthforward.  The distribution in Figure 4 of Risaliti et
al. (1999) is the expected one at the flux limit where most of the XRB
is resolved. Our survey resolves only a fraction of the XRB and
therefore it misses the sources in which large absorption
(log$N_H>24$) reduces the 5-10 keV flux below our flux limit.

\begin{figure}
\centerline{ \epsfig{figure=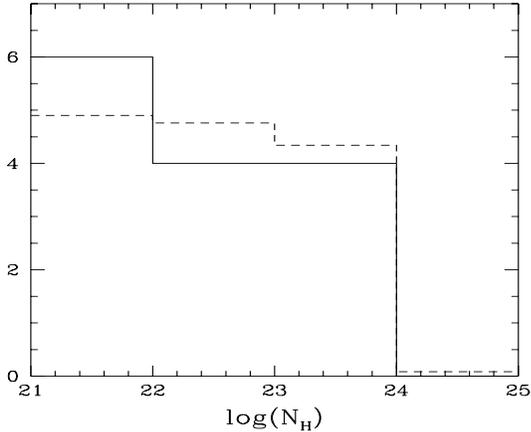, height=6.5cm, width=8.5cm,angle=0}
}
\caption[h]{
The column density distribution for the 14  HELLAS sources (solid line).
All upper limits have been collected in the log$N_H<22$ column.
The two objects with log$N_H=23\pm1$ appear in the column with
$23<logN_H<24$.
The dashed line represents the prediction of 
an upgraded version of the Comastri et al. (1995) syntesis
model in the 5-10 keV band and at the flux limit of
the sample presented in this letter ($8\times10^{-14}$ \cgs).
}
\end{figure}

The hard X-ray to optical ratio (X/O) of HELLAS AGN is similar to that
of nearby Seyfert 1 and Compton thin Seyfert 1.8-1.9 galaxies (from
the samples of Bassani et al. 1999, Matt et al 1999).  The HELLAS
sources can so be considered as their higher redshift (up to z=0.4)
analogs. It is worth noticing that another intermediate type 1.9,
Compton thin quasar has been discovered by Almaini et al (1995) and
Georgantopoulos et al. (1998) at even higher redshift (z=2.35).
Conversely, the X/O of the local Compton thick Seyfert 2
galaxies (Maiolino et al. 1998) with $F_{5-10keV}=2-10\times10^{-13}$
\cgs is lower by a factor 30-100.  The study of faint Compton thick
Seyfert 2 (Maiolino et al. 1998) at z=0.2--0.4 and V$\sim19$ (which
should have $F_{5-10keV}=10^{-15}-10^{-14}$ \cgs) must therefore await
instruments with sensitivity higher than ASCA and BeppoSAX, like those
on board AXAF and XMM.


\section{Discussion and Conclusions}

Our results, obtained using a purely hard X-ray selected sample and
based on new identifications as well as correlation of the HELLAS
sample with catalogs of known sources, provide strong evidence that
most of the sources resolved down to a 5-10 keV flux level of
$8\times10^{-14}$ \cgs are AGN. The amount of absorption in these AGN
is consistent with the prediction of AGN synthesis models of the XRB
(Comastri et al. 1995) in the 5-10 keV band and at the above flux
limit. However, the source breakdown is rather peculiar. In addition
to five `blue' continuum broad line quasars, which dominate optical
and soft X-ray surveys, we found 7 ``intermediate'' AGN. These objects
are ``intermediate'' both for their optical spectra which pose them
between type 1 and type 2 AGN and because of their X-ray absorption,
which is typically log$N_H$=22.5-23 (estimated from hardness
ratios). `Red' quasars have been selected in the past in the radio
band (Smith \& Spinrad 1980, Webster et al. 1995) or in soft X-rays
(provided that the absorbing column is not too large, A$_V\ls2$,
log$N_H<21.7$, Kim \& Elvis 1998). Their number density relative to
`blue' quasars is thought to be a few percent.  In contrast, we found
a fraction of $\sim$30\%, showing that hard X-ray selection may be a
more efficient and less biased way to search for `red' quasars.  The
fraction of AGN showing evidence of extinction/absorption to the
nucleus in optical/X-ray (7 out of 12 AGN) is $0.58^{+0.17}_{-0.18}$
(Gehrels 1986, $1\sigma$ confidence level).  The fraction of
intermediate AGN + narrow line galaxies in the ROSAT 0.5-2 keV survey
of the Lockman hole ($F_{0.5-2keV}>5.5\times10^{-15}$ \cgs, Schmidt et
al. 1998), and in the shallower Cambridge-Cambridge ROSAT Serendipity
Survey (CRSS, $F_{0.5-2keV}>2\times10^{-14}$ \cgs, Boyle et al,
1995a,b) is 0.28$\pm$0.09 (11 objects out 39) and 0.15$\pm$0.05 (12
objects out of 80), respectively.  The same fraction in the ASCA-ROSAT
combined surveys of Boyle et al.  (1998) and Akiyama et al. (1998) is
0.33$\pm$0.08 (20 objects out of 61).  We calculated, using both the
Fisher exact probability test (Siegel 1956) and the Barnes (1994) test
on the difference of two proportions, the probability that the HELLAS
and the Lockman hole, CRSS and ASCA-ROSAT samples differs in the
proportion of obscured objects.  This probability is $\sim95\%$ for
HELLAS/Lockman hole samples, $\sim99\%$ for the HELLAS/CRSS samples
and $\sim90\%$ for the HELLAS/ASCA-ROSAT samples, for both statistical
tests.  While more HELLAS identifications will clearly be useful to
strengthen this result, the above probabilities already indicate a
difference between the HELLAS sample and at least the two ROSAT
samples. Furthermore, while all HELLAS `blue' quasars in PSPC fields
are detected down to a typical WGACAT flux limit of
$1-5\times10^{-14}$ \cgs, only three of the six `intermediate' AGN in
PSPC fields are detected at this flux limit (see Table 1 and Sect 2),
again indicating that ROSAT surveys are more biased against faint
highly obscured objects that our survey.

We did not find any luminous ($L_X>10^{44}$ erg s$^{-1}$, highly
absorbed (log$N_H>23$) quasar~2.  However, one of the `red' quasar in
our sample has an high X-ray luminosity and may have an absorbing
column as high as log$N_H$=23.  Almaini et al. (1995) discovered an
highly obscured quasar at z=2.35, which has been classified as an
intermediate 1.9 AGN by Georgantopoulos et al. 1998.  Akiyama et
al. (1998) report the discovery of 2 luminous broad line quasars with
a very hard (possibly absorbed) X-ray spectrum.  One of our `blue'
quasar may have absorption in excess of log$N_H$=22.  Although the
statistics are not yet good enough to reach firm conclusions, these
few positive detections and the absence of even a single detection of
a narrow line quasar~2 (Halpern et al. 1998) suggest that X-ray
absorption in high luminosity objects may be associated more
frequantly with intermediate type 1.5-1.9 AGN, `red' quasars, or even
normal `blue' quasars, rather than with pure type 2 objects.  The bulk
of the hard XRB could be made by a large population of Seyfert 1.8--2
like galaxies and moderately absorbed `red' and intermediate quasars
(see also Kim \& Elvis 1998).  This is in line with scenarios where
the absorption takes place in a starburst region surrounding the
nucleus (Fabian et al. 1998). In this case the amount of the absorbing
gas may depend on the nuclear mass (and luminosity), and then the
evolution of the type 2 AGN luminosity function would also be strongly
affected by the nuclear environment and may imply a link between AGN
evolution and the history of star-formation as measured from optical
and sub-mm surveys (see e.g. Madau et al. 1996, Lilly et al. 1998). In
this regard, it is worth remarking that a sizeable population of low
luminosity AGN (M$_B$ in the range --17; --20) may exist at very faint
optical fluxes (V=26-27) in the Hubble Deep Field North (Jarvis \&
MacAlpine 1998).  This conclusion will soon be tested by the deep and
high spatial resolution surveys that will be performed by AXAF and XMM
in the next few years. These deep surveys, together with the shallower
but larger area surveys perfomed by ASCA and BeppoSAX, should be able
to discriminate between different scenarios for the XRB and provide
information on the connection between AGN and galaxy evolution.

\bigskip
\centerline{\bf Acknowledgements}

We thank the BeppoSAX SDC, SOC and OCC teams for the successful
operation of the satellite and preliminary data reduction and
screaning, A. Matteuzzi for his work on
MECS source position reconstruction, P. Massey for his help at the 
Kitt Peak 4m telescope, P. Berlind for the Whipple 60'' spectra of the 
1SAXJ1519.5+6535 field, G. C. Perola, M. Vietri and G. Zamorani  
for a careful reading of the manuscript and useful discussions.

\end{document}